# Direct single-shot imaging of time-resolved terahertz vector fields

J. Lafrenière-Greig[1], J. E. Nkeck[1], X. Ropagnol[1,3], G. Gandubert[1], Salim Hmidi[1], O. Brincoveanu[3], O.-G.Simionescu[3], E. Abraham[4], R. Tudor[3,†] and F. Blanchard[1,*]
[1]Département de Génie Électrique, École de technologie supérieure (ÉTS), Montréal, Québec, Canada
[2]INRS – EMT Institut national de recherche scientifique, Varennes, Québec, Canada
[3]National Institute for Research and Development in Microtechnologies IMT, 077190 Bucharest, Romania
[4]Université de Bordeaux, CNRS, Laboratoire Ondes et Matière d'Aquitaine (LOMA), UMR 5798, Talence, France
†rebeca.tudor@imt.ro , *francois.blanchard@etsmtl.ca

**Abstract**

We demonstrate direct single-shot retrieval of the full in-plane vector field of broadband terahertz waves using electro-optic sampling. Implemented with a circularly polarized optical probe and polarization-sensitive camera, the method overcomes the projection constraints of conventional electro-optic sampling and enables direct time-domain reconstruction of transverse vector fields without Fourier-domain inversion or sequential polarization analysis. We experimentally validate the approach through spatiotemporal imaging of linear, circular, azimuthal and radial terahertz fields generated by polarization-engineered silicon metamaterials. The measurements reveal ultrafast time-resolved vector-field dynamics inaccessible to conventional scalar detection, including helicoidal field rotation within a single optical cycle and the spatiotemporal evolution of structured polarization topologies. Beyond structured-field imaging, we demonstrate single-scan vector-field spectroscopy through measurements of quartz birefringence and broadband terahertz waveplate performance. More broadly, this work transforms electro-optic sampling from a projection-based measurement into a vector-resolved imaging and spectroscopy platform for structured electromagnetic fields.

## Introduction

Electromagnetic waves are intrinsically vectorial, with spatial phase gradients and polarization textures governing local momentum, topology, and energy flow [1, 2, 3]. As structured terahertz (THz) radiation increasingly incorporates orbital angular momentum, polarization singularities, and spatiotemporal vector structure, scalar or intensity-based descriptions become insufficient for capturing the full organization of the field [1, 3, 4, 5]. Yet despite rapid advances in structured THz generation and wavefront engineering [7], direct experimental access to the complete time-resolved vector electric field remains limited.

Since its introduction in 1995 [8], electro-optic sampling (EOS) has served as the central technique for broadband, phase-resolved detection of THz electric fields, underpinning applications ranging from ultrafast spectroscopy [9] to coherent imaging [10] and emerging quantum field sensing schemes [11]. However, conventional EOS fundamentally operates as a projection measurement: a linearly polarized optical probe samples only a single component of the THz electric field through tensor-mediated birefringence in the detection crystal [6, 9, 10, 11]. Accessing orthogonal field components therefore requires sequential probe rotations, crystal rotations, or multi-step polarization analysis [12, 13, 14], preventing direct and simultaneous retrieval of the complete vector field. Even in full-field implementations, the measured signal generally contains mixed contributions from the transverse components, while the anisotropic electro-optic response of commonly used zinc-blende crystals introduces orientation-dependent sensitivity and trade-offs between efficiency and completeness [11]. As a result, real-time vector-field imaging has remained experimentally complex and poorly suited to rapidly evolving or non-repetitive electromagnetic dynamics.

Here we establish a generalized tensor-resolved electro-optic formalism that enables direct, single-acquisition retrieval of the in-plane THz vector electric field in the time domain. Implemented using a circularly polarized optical probe in a (110)-cut zinc-blende crystal, the method decouples field amplitude and orientation, overcoming the projection constraints of conventional linear-probe EOS while preserving broadband operation. We validate the formalism experimentally using polarization-engineered THz metamaterials designed to generate structured vector fields with spatially varying polarization topology and orbital angular momentum [17]. The reconstructed measurements directly resolve the instantaneous organization and evolution of the transverse electric-field vector, revealing spatiotemporal polarization dynamics that remain inaccessible to scalar, projection-based, or sequential detection schemes. In quantitative agreement with electromagnetic theory, this integrated scheme is transferable to other electro-optic materials and near-field geometries, providing a simplified route toward vector-resolved metrology of structured and nonclassical THz fields.

## Results

**Single-acquisition THz vector-field imaging scheme.** To overcome the limitations of conventional projection-based electro-optic sampling, we develop a detection scheme that enables direct, single-acquisition retrieval of the transverse vector fields (TVFs) of broadband THz transients in the time-domain. The method requires only the two camera exposures associated with dynamic probe subtraction [18], while providing complete in-plane vector information without sequential measurements.

The key advance lies in combining a circularly polarized probe with a polarization-sensitive camera (PSC), enabling simultaneous acquisition of the full polarization state in a single exposure. In contrast to polarization-resolved EOS implementations based on linear probes and selected analyzer channels (e.g., 0°/90° projections) [19], which remain inherently projection-based, our approach distributes the probe across a four-state linear polarization basis ($I_{0°}$, $I_{45°}$, $I_{90°}$, $I_{135°}$) in parallel (Fig. 1a, right inset), following Stokes polarimetry principles [16].

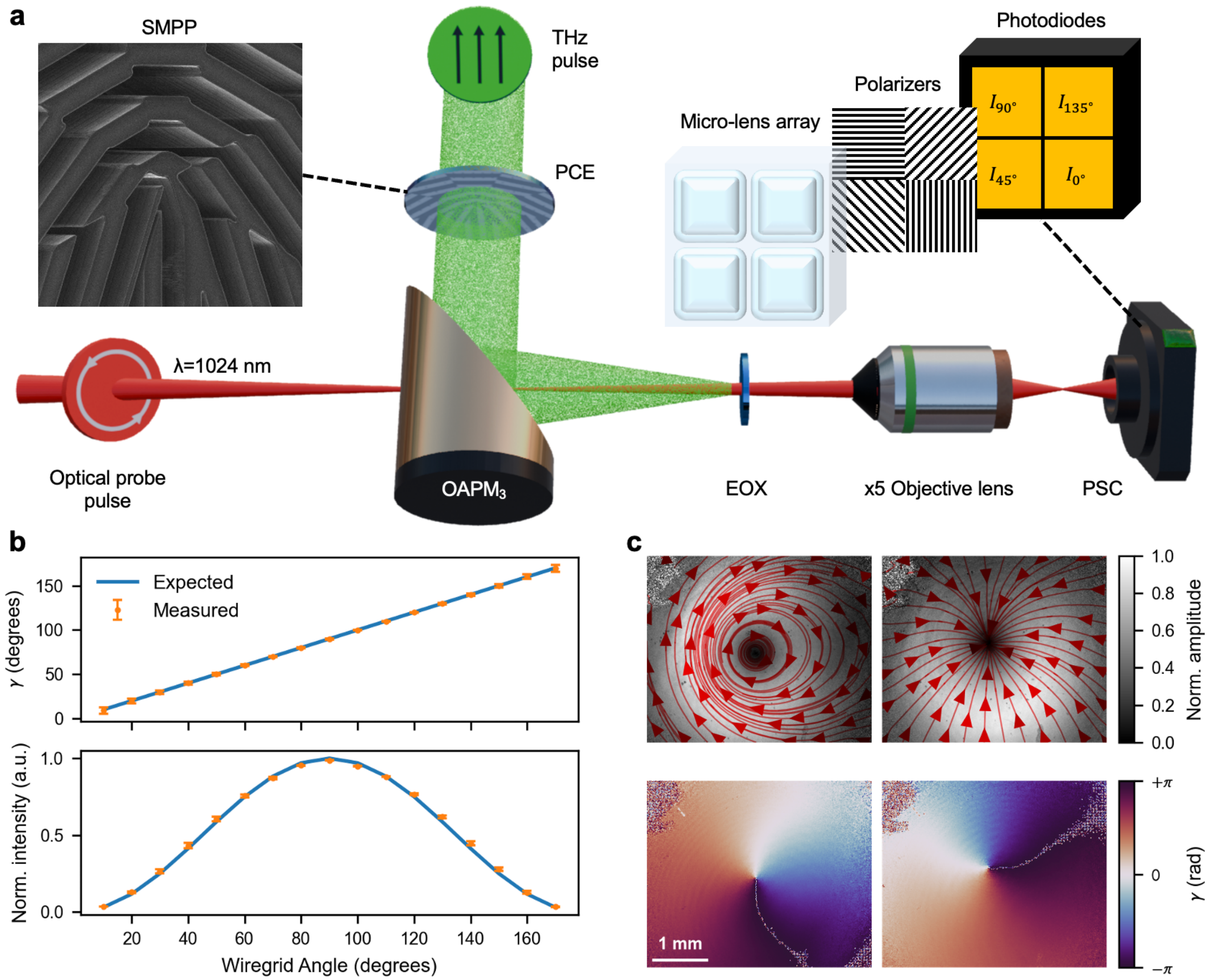

**Figure 1 | Electro-optic configuration for single-acquisition vector field imaging. a**, Experimental setup. A polarization control element (PCE), such as a wire-grid polarizer (WGP), segmented metasurface phase plate (SMPP), or achromatic metamaterial wave plate (AMWP), structures the THz field, which is focused onto an electro-optic crystal (EOX) by an off-axis parabolic mirror (OAPM3). A circularly polarized probe pulse (λ = 1024 nm) co-propagates through the EOX. A 5x objective images the crystal onto a polarization-sensitive camera (PSC), which records THz-induced polarization modulation. Insets: SMPP and PSC superpixel with analyzers at 0°, 45°, 90°, and 135°. **b**, Validation with a rotating WGP: retrieved polarization angle (top) and normalized intensity following Malus' law (bottom). **c**, Cylindrical vector beams generated by the SMPP. Top: normalized electric-field amplitude with transverse vector field lines (red). Bottom: orientation maps γ for azimuthal (left) and radial (right) states.

The circular probe ensures equal optical power along the principal axes of the electro-optic crystal, independent of the incident THz field orientation. This balanced interaction mitigates the anisotropic response imposed by the electro-optic tensor in (110)-cut zinc-blende crystals [13], enabling analytical retrieval of the in-plane electric-field amplitude |E(x,y,t)| and orientation γ(x,y,t) in the laboratory frame without mechanical rotation or polarization beam splitting (Methods). The acquisition rate is thus limited only by pump modulation and camera frame rate.

We validate the quantitative accuracy of the measurement by recording normalized Stokes parameter maps ($s_1$, $s_2$) at the pulse peak for different WGP orientations. Calibration of the laboratory reference angle β (Methods) yields the dependencies shown in Fig. 1b. The retrieved γ angle follows the WGP orientation with R2=0.9999 and a maximum sensitivity of 0.418° (Methods) (Fig. 1b, top), while $|E|^2$ obeys Malus' law with $R^2$ = 0.997 (Fig. 1b, bottom). These results confirm that the formalism decouples amplitude and orientation in the (110)-cut zincblende crystal. It should be emphasized that the sensitivity remains anisotropic due to the crystal geometry. A field oriented along the crystal's [-1,1,0] axis induces a birefringence variation that is double that of a field of the same magnitude oriented along the [0,0,1] axis [18, 19].

To demonstrate real-time vector imaging, we measure the TVFs of cylindrical vector beams (CVBs) generated by a 3D-engraved silicon segmented metamaterial phase plate (SMPP) (Fig. 1a, left inset) [22]. Such beams exhibit spatially varying polarization topology characteristic of structured light fields [23]. With the delay line fixed at the THz peak, azimuthal and radial states are immediately distinguished through their TVF lines reconstructed from $|E|$ and γ (Methods) (Fig. 1c). Although both beams exhibit identical doughnut-shaped amplitude profiles, their polarization topology differs fundamentally with fields following either the azimuthal ($\mathbf{E} \propto \hat{\phi}$) or radial ($\mathbf{E} \propto \hat{r}$) directions. The two configurations are related by a local π/2 rotation of the polarization vector at each spatial position, which is directly resolved in the γ maps.

**Spatiotemporal topology of canonical states.** By scanning the probe delay (Methods), we record time-resolved transverse vector-field (TVF) dynamics at the crystal plane, analogous to standard time-domain field reconstruction in THz time-domain spectroscopy (THz-TDS) [6, 9]. Figure 2a–d shows representative TVF snapshots at t = −0.5, 0, and +0.5 ps relative to the pulse peak, together with the corresponding γ maps (full spatiotemporal videos available in supplementary V1-4). For the linearly polarized beam (Fig. 2a), the reconstructed vectors remain aligned across the Gaussian amplitude profile, with γ spatially uniform within the high-signal region. The observed direction inversions of the fields follow the polarity reversal of the THz transient, confirming faithful time-domain measurement of the instantaneous transverse electric field.

In contrast, beams generated by the SMPP (Fig. 2b,c) exhibit a temporal drift of the central singularity between −0.5 and +0.5 ps. This behavior arises from chirp introduced by group-velocity dispersion within the THz generation crystal [24]. Because lower frequencies dominate the leading edge of the pulse and higher frequencies dominate the trailing edge, the relative phase between the vortex ($LG_{0,\pm1}$) and fundamental Gaussian ($LG_{0,0}$) modes evolves continuously. The resulting modal interference shifts the singularity position across the beam profile, consistent with the propagation and interference properties of LG modes carrying orbital angular momentum [23, 24] and with the spectral modal structure discussed in Fig. 3.

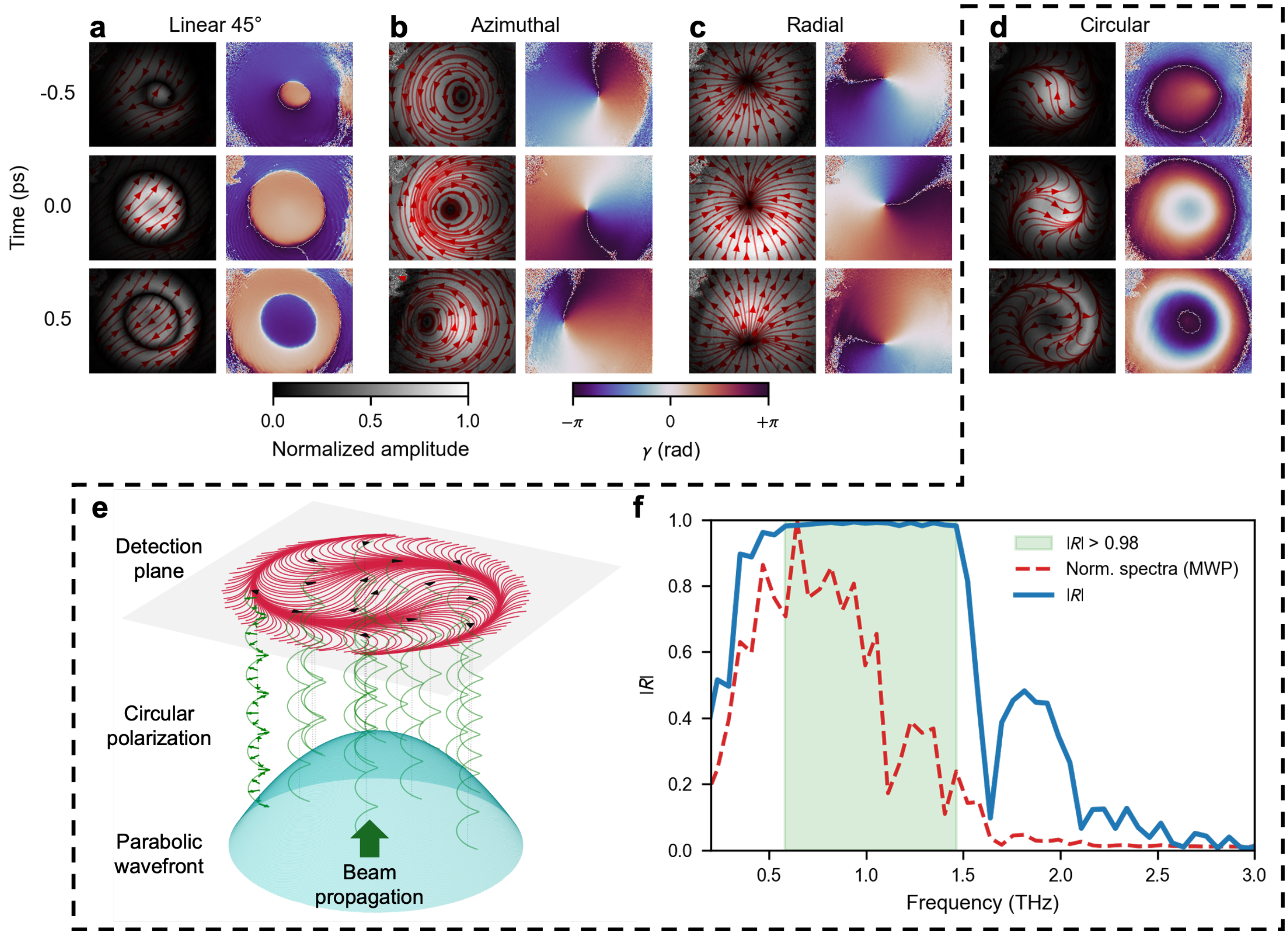

**Figure 2 | Spatiotemporal topology and phase-polarization mapping. a–d**, Time-resolved transverse vector fields for linear (45°) (a), azimuthal (b), radial (c), and circular (d) polarization states. Rows correspond to t = −0.5, 0, +0.5 ps relative to the pulse peak. Left panels: normalized electric-field amplitude with instantaneous field lines (red). Right panels: corresponding orientation γ maps. **e**, Schematic of phase–polarization coupling for a circularly polarized beam, illustrating mapping of wavefront curvature onto the detection plane. **f**, Correlation between the spatially unwrapped phase in the frequency domain and γ at the pulse peak timing. Blue: absolute correlation coefficient |R|. red dashed: normalized amplitude spectrum of the beam generated by the AMWP. Shaded region: bandwidth where |R| ≥ 0.98.

To investigate the spatiotemporal dynamics of circularly polarized THz waves, we employ an achromatic metamaterial waveplate (AMWP) engineered to provide broadband $\pi/2$ phase retardation (Materials). The reconstructed TVF traces a helicoidal trajectory in time (Fig. 2d), directly resolving the rotation of the transverse electric-field vector within a single optical cycle. The measured rotation rate of 0.768 THz matches the central frequency of the transmitted wavepacket (Fig. S1), consistent with the expected temporal evolution of broadband circularly polarized fields [17]. The vector maps additionally reveal spiralling spatial patterns arising from coupling between the polarization rotation introduced by the AMWP and the parabolic wavefront curvature imposed by the OAPM. Because the local phase gradient $\nabla\phi$ governs the local wavevector and momentum direction in structured electromagnetic fields [1], rays at larger radii accumulate larger effective phase delays during propagation, producing a radial gradient in $\gamma$. As illustrated schematically in Fig. 2e, the wavefront curvature is therefore directly encoded into the polarization orientation measured at the detection plane.

This phase–polarization coupling reflects known spin–orbit and geometric-phase interactions in structured light [1, 2, 26]. To quantify this relationship, we compute the absolute correlation coefficient $|R|$ between the spatially unwrapped spectral phase $\phi(x, y, \nu)$, obtained from Fourier analysis of the full delay scan, and the $\gamma$ map at $t = 0$ ps. The correlation is evaluated within the beam FWHM to avoid orientation-wrapping discontinuities. As shown in Fig. 2f, $|R|$ exceeds 0.98 over 0.6–1.45 THz, demonstrating broadband correspondence between spatial phase topology and instantaneous polarization orientation. These measurements therefore provide direct experimental access to the evolving spatiotemporal organization of structured electromagnetic fields in the time domain.

**Modal decomposition of cylindrical vector beams.** Fourier transformation of sampled time-domain image stack yields polarization-resolved spectral amplitude and phase (Fig. 3a), following standard THz-TDS procedures [28]. In the circular polarization basis, the field can be expressed as a superposition of orbital angular momentum (OAM) modes with well-defined helicity [1, 27]. This representation highlights the coupling between spin (polarization) and orbital angular momentum in structured electromagnetic fields, forming a classical analogue of quantum-like nonseparability due to the intrinsic coupling between polarization and spatial degrees of freedom [28, 29, 30]. To experimentally quantify this coupled structure, we project the measured complex field onto a Laguerre-Gaussian (LG) basis, which forms a complete set of solutions to the paraxial wave equation carrying quantized OAM [33]. Using normalized overlap integrals between the field and these basis functions [32, 33, 34], our analysis reveals dominance of the $LG_{0,\pm1}$ vortex mode with ≥ 50% purity across a 0.31 THz bandwidth (0.77–1.08 THz) (Fig. 3b). A peak purity of the $LG_{\mathbf{0,\pm1}}$ mode of 0.73 at 0.88 THz is measured in agreement with the design frequency of the device (Materials).

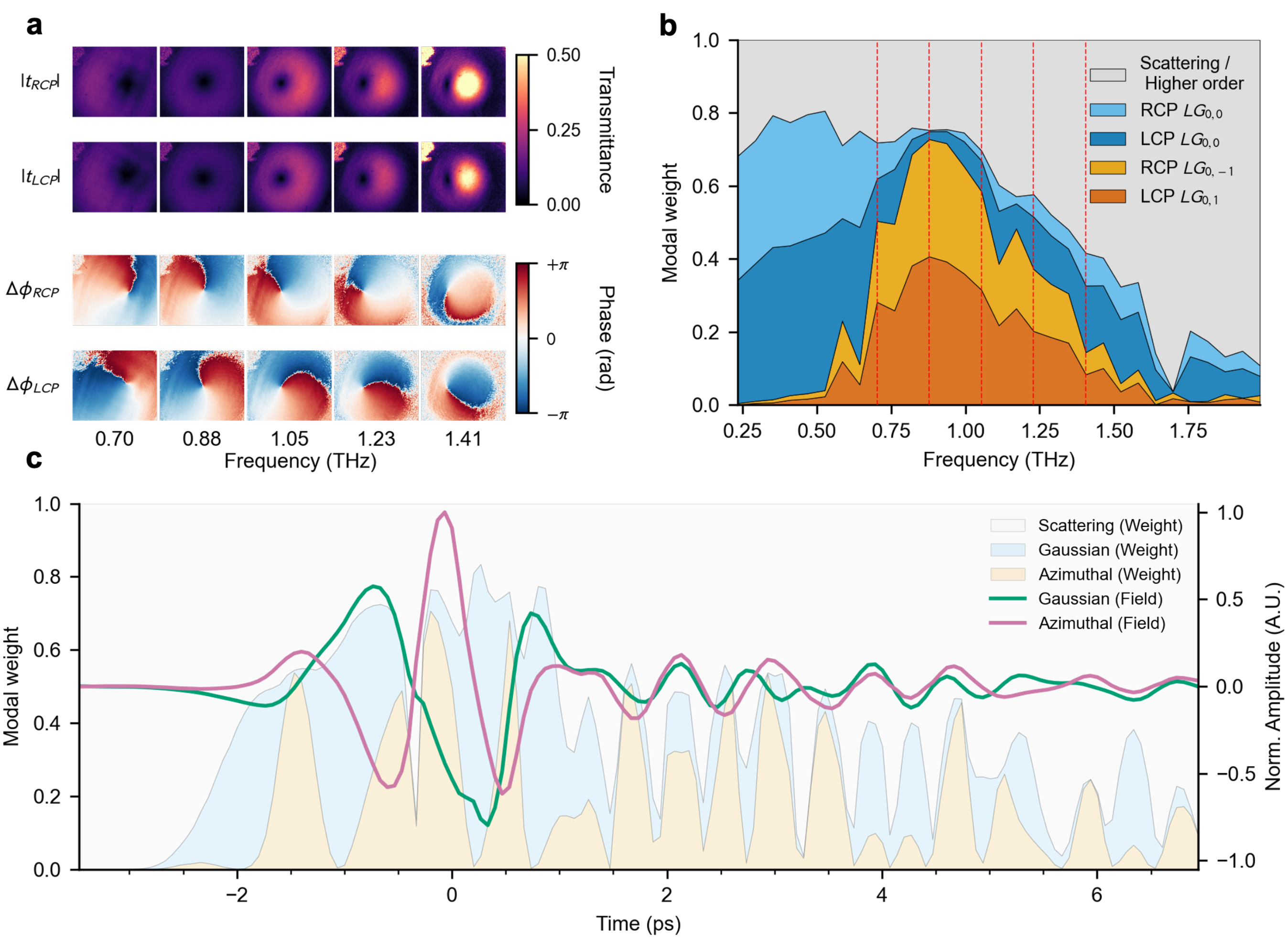

**Figure 3 | Full characterization of the THz cylindrical vector beam**. **a,** Reconstructed spectral transmittance |t| and phase Δϕ for the right- and left-circular polarization components (RCP, LCP) at selected frequencies from 0.70 to 1.41 THz. **b,** Modal weight across the pulse bandwidth obtained by projection onto the LG basis. Stacked areas indicate modal weights for RCP and LCP components. **c** Time-domain CVB modal weights obtained with the same projection. Scattering and higher order modes are combined in the grey region. The normalized fields of both propagation modes are overlaid on the plot.

Importantly, this modal purity is extracted directly from the single-acquisition vector-field reconstruction, without sequential polarization projections or interferometric references typically required for OAM analysis [35, 36]. Because the reconstruction provides the full complex vector field per helicity channel, the decomposition captures the frequency-dependent redistribution between $LG_{0,\pm1}$ and $LG_{0,0}$ components. This redistribution quantitatively explains the spatiotemporal drift of the beam singularity observed in Fig. 2b,c, linking temporal topology to spectral modal content, consistent with interference between co-propagating OAM modes [1, 24]. Toward the spectral edges, increasing Gaussian and higher-order contributions reduce modal purity, consistent with metasurface dispersion and simulated response (Supplementary Fig. S5).

Finally, we directly project the measured time-resolved electric fields onto the LG vector basis at each temporal step to extract the ultrafast topological propagation dynamics of the THz wavepacket generated by the SMPP. The ability of our system to simultaneously capture the transverse vector fields ensures that the measured time delays are linked only to the generation and propagation dynamics of the beam.

The modal transient field traces (Fig. 3c) reveal the spatiotemporal coupling between the Gaussian and azimuthal components of the THz pulse. This correlation between the beam's spatial and temporal degrees of freedom manifests as a relative temporal walk-off and complex interference around the main peak (-2 ps to 2 ps) between the co-propagating modes [37, 38]. Although the Gaussian component arrives first (t=-1.70 ps), it is the wavepacket's azimuthal mode that peaks first (t=0.0 ps). This temporal walk-off stems from effects such as differing Gouy phase shifts [22] and group delay dispersion within the SMPP [40]. We also observe a π phase shift between the modes, introduced by the spatially arranged half-waveplates operating within the SMPP's bandwidth.

Remarkably, the transient azimuthal mode achieves a maximum modal purity of 0.71 near the peak ($t = -0.2$ ps), closely matching the limit extracted from the frequency-domain analysis. We also observe a timing mismatch between the peak modal purity and field of the generated CVB , which we attribute to the strong residual presence of the Gaussian mode. Identifying this is crucial, as it enables the tailoring of the properties of pulse-shaping samples to ensure that the generated mode's purity and peak field align. Ultimately, this time-resolved vectorial modal analysis provides a direct window into how dispersion and frequency-dependent diffraction dynamically reshape structured THz wavepackets.

**Single-scan vector-field spectroscopy and polarimetry.** Beyond mapping complex topological propagation, the single-scan vector field retrieval functions as a complete platform for broadband polarimetric spectroscopy. To demonstrate this capability for advanced material and device characterization, we insert a 200 μm thick quartz sample or the AMWP before the detector, aligning the detection frame to the material axes in post-processing (Fig. 4a). We extract the slow and fast refractive indices from a single delay scan by orienting the sample 45° relative to the incident linearly polarized THz field. Using standard phase analysis (methods), we measure a birefringence $(\Delta n = n_s - n_f)$ of 0.05 at 1 THz, in excellent agreement with the literature [26, 39, 40] (Fig. 4b).

Applying this same single-scan framework to complex metamaterials, the AMWP is characterized in terms of its transmittance ($|t|$), retardance ($\Delta\phi = \phi_s - \phi_f$), and Stokes parameters ($s_1$, $s_2, s_3$) (Fig. 4c). This rapid characterization reveals deviations from ideal $\pi/2$ retardance at the spectral extremes. Defining the operational window by $s_3 > 0.95$ yields a 0.72 THz bandwidth (shaded region in Fig. 4c, bottom plot), with an average $s_3$ of 0.99 (polarization conversion ratio (PCR) of 99.8%). Using this system, we efficiently determine that the AMWP converts 32% of incident linearly polarized THz to circularly polarized THz, accounting for total transmission (Methods).

Collectively, these measurements demonstrate that our unified detection architecture not only provides quantitative reconstruction of the complete time-domain transverse vector field and its spectral modal content, but also establishes a robust, highly efficient metrology platform for evaluating next-generation THz pulse-shaping devices and anisotropic materials.

## Discussion

Conventional EOS provides broadband, phase-resolved detection of THz fields but intrinsically measures only a single projection determined by probe polarization and crystal tensor symmetry [6, 8]. Access to orthogonal components generally requires sequential rotation or multi-step analysis, breaking simultaneity for dynamic structured fields. The present architecture preserves the broadband sensitivity of EOS while enabling single-acquisition retrieval of the in-plane vector field, mitigating projection-induced ambiguities without interferometric complexity.

Full time-domain vector reconstruction provides direct access to the coupled amplitude–phase–orientation dynamics of structured THz radiation. The observed singularity drift and helicoidal rotation illustrate how dispersion and modal interference dynamically reshape the instantaneous topology of structured wavepackets, phenomena typically inferred indirectly from spectral measurements [1, 23, 24]. Because the complete complex vector field is recovered, OAM content can be decomposed numerically without mode-specific phase elements or inverted staircase holograms [35, 36]. The work further enables direct time-resolved modal decomposition of structured THz wavepackets, revealing ultrafast evolution of OAM content and spatiotemporal topology. Arbitrary superpositions of helicity-resolved modes can therefore be retrieved in post-processing, providing a flexible framework for broadband OAM multiplexing in the THz regime and for probing complex spin-orbit interactions such as the transfer of OAM to plasmonic excitations [43]. Vector-resolved acquisition further simplifies analysis of anisotropic and dispersive samples by decoupling the detection frame from the material axes, eliminating mechanical basis rotation [12, 13] and reducing reliance on large-scale broadband simulations.

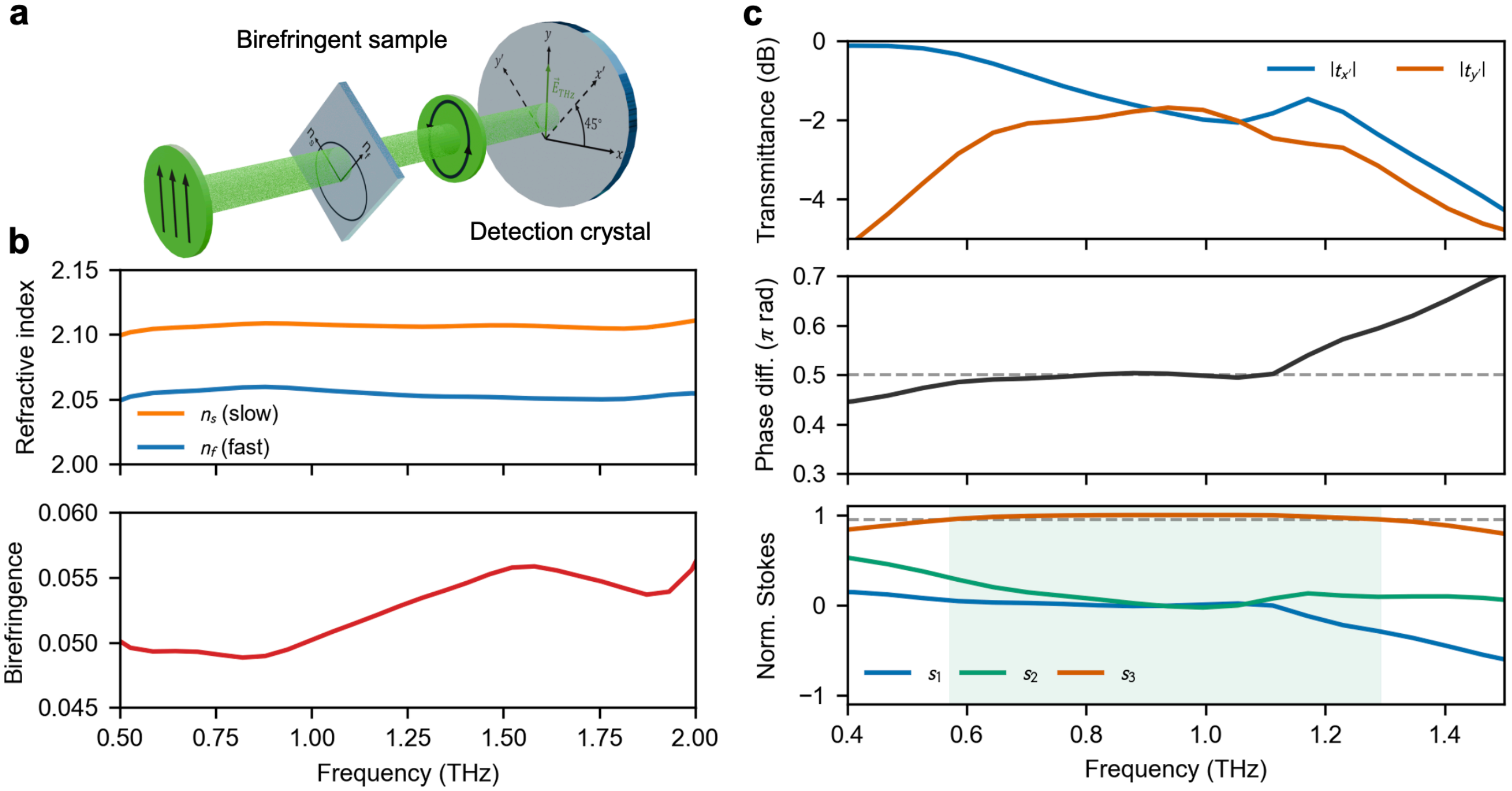

**Figure 4 | Quartz and AMWP characterization through single-scan vector field spectroscopy. a,** Measurement geometry. A birefringent sample is placed before the detection crystal; the detection basis is numerically rotated in post-processing to align with the slow ($n_s$) and fast ($n_f$) axes. **b,** Quartz characterization from a single delay-line scan. Top: frequency-dependent refractive indices along the slow and fast axes. Bottom: extracted birefringence. **c, AMWP** performance. Top: transmission spectra $|t_{x'}|$ and $|t_{y'}|$. Middle: phase retardation between orthogonal components. Bottom: normalized Stokes parameters ($s_1$, $s_2$, $s_3$) of the transmitted THz field. Shaded region indicates the bandwidth where $s_3 > 0.95$.

Beyond probing light-matter interactions, this architecture provides a critical diagnostic platform for the rapidly expanding field of THz flat optics [44]. As demonstrated by our characterization of the SMPP, the ability to map time-resolved vector fields exposes complex spatiotemporal dynamics that are entirely obscured by standard frequency-domain or time-integrated analyses. Identifying these transient imperfections, including the timing mismatch between maximum modal purity and peak field amplitude, provides the direct experimental feedback required to tailor the geometric dispersion of future metasurfaces. This capability will be essential for the iterative design of advanced pulse-shaping components in the THz regime.

Despite the formalism mathematically correcting for the amplitude dependence on $\gamma$, the detection sensitivity remains inherently anisotropic [13]. While our intense THz source largely masks this constraint, a potentially superior future configuration would utilize (111)-cut zincblende crystals to achieve transverse isotropic sensitivity [19, 42, 43]. This shift would incur an 18% reduction in maximum THz-induced modulation compared to the (110) geometry, but avoids the 50% sensitivity drop the current configuration suffers when the THz field aligns with its weak [0,0,1] axis. Beyond crystal geometry, the precision of the $\gamma$ mapping is limited by the near-infrared extinction ratio of the integrated polarizers and the quantum efficiency of the sensor. Shifting the probe to the visible regime (e.g., 512 nm) would yield an order-of-magnitude improvement in both metrics, provided detection materials like ZnS are employed to preserve phase matching [47]. Furthermore, the system's operational bandwidth remains fundamentally constrained by phonon absorption and phase-matching limits within the electro-optic crystal [4, 29]. Exploring materials with broader phase-matching windows, such as GaP [49], offers a clear pathway to extend the spectral characterization window, albeit with a necessary trade-off in absolute detection sensitivity.

Given the static crystal, the formalism is compatible with near-field electro-optic imaging [10, 17] while also suggesting routes toward polarization-resolved THz topography [50] and real-time decoding of generalized CVBs [51]. More broadly, single-scan vector-field 2D-EOS bridges ultrafast spectroscopy with structured and topological optics, providing direct experimental access to the intrinsically vectorial organization of broadband electromagnetic fields [39].

More broadly, vector-resolved EOS provides experimental access to the instantaneous organization of broadband electromagnetic fields in space and time. This capability opens a route toward direct studies of ultrafast topological dynamics, structured vacuum fluctuations, spin-orbit interactions, and nonseparable electromagnetic states beyond the limits of projection-based detection.

## Methods

**Experimental setup.** The detailed optical layout is provided in Supplementary Section S1. The system employs a PHAROS Yb-based femtosecond laser from LIGHT CONVERSION (1024 nm wavelength, 280 fs pulse duration, 10 W average power, and 25 kHz repetition rate), following our previously reported implementation [52]. The pump beam (90%) was directed to the generation stage, while the remaining 10% served as the probe for electro-optic sampling (EOS). THz radiation was generated using the tilted pulse-front scheme in a lithium niobate (LN) crystal [52]. It was then collected and collimated using off-axis reflective optics and focused onto a 1 mm thick CdTe crystal. CdTe was chosen for favorable phase matching at the 1.024 μm probe wavelength [24]. The THz beam interacted with a left-circularly polarized probe prepared using a polarizer and quarter-wave plate. The probe pulse was temporally compressed to ~50 fs to improve the sampling limit.

The transmitted probe was imaged onto a polarization-sensitive camera (Alkeria Celera-P series, Sony IMX250MZR sensor) featuring division-of-focal-plane linear polarizers integrated at the superpixel level. A 5x microscope objective provided a magnification of 1.89, yielding an optical resolution of 3.65 μm per superpixel at the crystal plane. Images were acquired with 12-bit depth at 50 fps with a 10 ms exposure time. Dynamic background subtraction was performed by modulating the THz pump beam with a mechanical chopper synchronized to half the camera frame rate. For each delay position, 100 pump-on and 100 pump-off frames were averaged, yielding a dynamic range of 58.7 dB per transverse component under a 45° wire-grid polarizer reference. Time-domain vector-field traces were obtained by scanning the optical delay line in 66.7 fs steps. Spectral information was retrieved via fast Fourier transformation (FFT) of the time-domain traces. Spatial binning (4x4) was applied during time-domain scans to reduce data volume without affecting modal analysis.

**TVF detection and calibration scheme.** The transverse vector field is resolved by mapping the birefringence in the (110)-cut zinc blende crystal induced by a field having an arbitrary orientation $\boldsymbol{\alpha}$ in the transverse plane [18, 19]. We employ a Jones matrix formalism to link the probe's polarization state to the THz induced phase shift $\boldsymbol{\Gamma}$ and the rotation $\boldsymbol{\psi}$ of the principal axes (full derivations are provided in Supplementary Section S3) [12, 18]. Normalized Stokes parameters $\boldsymbol{s_1}$ and $\boldsymbol{s_2}$ are extracted from the camera's orthogonal polarizer pairs, and dynamic subtraction isolates the electro-optic contributions $\boldsymbol{\Delta s_1}$ and $\boldsymbol{\Delta s_2}$. The instantaneous orientation angle $\boldsymbol{\gamma}$ of the THz electric field in the laboratory frame is calculated as:

$$\gamma = \beta \pm \tan^{-1}\left(2\cot\left(\tan^{-1}\left(\frac{\Delta s_1}{\Delta s_2}\right) + 2\beta\right)\right) + n\pi, n \in \mathbb{Z}, \tag{1}$$

where β is the crystal orientation angle relative to the laboratory frame. The electric field amplitude $|E|$ is extracted by correcting for the resoponse of the (110) geometry according to:

$$|\mathbf{E}| \propto \sqrt{\frac{\Delta s_1^2 + \Delta s_2^2}{1 + 3\cos^2(\gamma - \beta)}}. \tag{2}$$

To apply these equations accurately, β is calibrated using a WGP and minimizing the mean squared error between the measured and reference electric field vectors expected from Malus law. Spatial maps of the calibrated crystal orientation and detailed anisotropic sensitivity analysis are available in Supplementary Sections S4 and S5. Streamlines were generated via 2nd-order Runge-Kutta integration, with directional arrows mapped to the midpoints of the calculated trajectories [53].

**Spectral analysis and modal decomposition.** Spectral amplitude and phase are retrieved by applying a FFT to the time domain image stack. The resulting hyperspectral data is normalized using either the transmission spectrum of a 525 μm thick high-resistivity silicon substrate or free propagation in air. To analyze the composition of the cylindrical vector beams, the transverse fields are converted to a circular polarization basis [23, 24]. The reconstructed electric fields $\boldsymbol{E_{rec}}$, on the left and right circular polarization channels, are modelled as a superposition of Laguerre Gaussian modes $\boldsymbol{LG_{p,l}}$:

$$E_{rec}(r,\phi,\nu) = \sum_{p,l} C_{p,l} LG_{p,l}(r,\phi), \tag{3}$$

where the complex coefficients $C_{\rho,l}$ represent the amplitude and phase of each mode. The chosen basis spans radial orders $\rho \in [0,3]$ and azimuthal orders $l \in [0, \pm 3]$. These coefficients are obtained by projecting the measured field onto the analytical modes. The same basis is used for the time domain vector beam decomposition by considering the γ map as analogous to the phase front (Supplementary Section). Despite the paraxial assumption, this projection leads to minimal spatial residual error between measured and reconstructed beams while avoiding overfitting artifacts (Fig. S4). Detailed equations for the overlap integrals and modal purity calculations are provided in Supplementary Section S6. Because projection onto the circular polarization basis maps both the vertical linear and azimuthal vector fields to a phase shift, these states can be exclusively isolated by extracting the imaginary part of the expansion coefficients, $\mathrm{Im}\{C_{p,l}\}$. These isolated components were subsequently used to plot the time-evolution of the modal fields.

For birefringence measurements, orthogonal components were acquired simultaneously in a single THz-TDS scan and normalized by the reference taken in air or in a 525 μm thick silicon wafer. A tapered flat-top temporal window was applied to suppress internal reflections in the 200 μm quartz sample (window parameters detailed in Supplementary Section S7). Orthogonal frequency (ν) dependent refractive indices were extracted from the reference normalized phase $\Delta\phi(\nu)$ using conventional phase unwrapping procedures [28]:

$$n(\nu) = 1 + \frac{\Delta\phi(\nu)c}{2\pi\nu d}, \quad (4)$$

with sample thickness $d$ and speed of light in vacuum $c$. The $s_3$ Stokes parameter was used to evaluate the AMWP performance:

$$s_3 = \frac{2E_s E_f \sin\left(\Delta\phi_s - \Delta\phi_f\right)}{E_s^2 + E_f^2}, \quad (5)$$

with electric fields $E_s$,$E_f$ and normalized phases $\Delta\phi_s, \Delta\phi_f$ along the slow and fast propagation axes respectively.

## Materials

**Segmented metasurface phase plate.** The segmented metasurface phase plate (SMPP) was fabricated at the National Institute for Research and Development in Microtechnologies (IMT). The SMPP, shown in the Fig. 1a inset, contains eight distinct sectors fabricated through dry etching. The micropatterns have a depth of 185 μm and a period of 30 μm, analogous to the four-sector SMPP . Fabrication was performed on a 525 μm thick high-resistivity (20 kΩ·cm) substrate. After substrate cleaning, a thermal treatment at 150 °C for 1 minute was performed using a Hexamethyldisilane (HDMS) adhesion promoter. In the next step, the positive photoresist AZ4562 was spin-coated to a thickness of 7 μm. After mask alignment and exposure for 38 seconds, the photoresist was developed, followed by deep reactive ion etching (DRIE). The photoresist was then removed, and the wafer was cleaned and diced into chips.

**Achromatic metamaterial wave plate.** The AMWP is a dielectric-metal hybrid metadevice also fabricated at IMT [17]. On one side, it contains an etched dielectric metasurface with a feature width of 12.5 μm and a period of 50.9 μm. The opposite side comprises subwavelength structures with a thickness of 100 nm, a width of 13.8 μm and a period of 101.6 μm. The detailed design and optimization of these novel metamaterials will be the subject of a subsequent publication.

**Acknowledgements**

This work was financially supported by the Natural Sciences and Engineering Research Council of Canada (NSERC, Grant No. 2023-03322) and the Canada Research Chairs Program (CRC-2024-00354), the Alliance International ALLRP590951-23, the Alliance Quantum grants (ALLRP 597331-24), and the Horizon Europe project Mimosa.